\documentclass[a4paper,11pt]{article}
\usepackage{pos}
\usepackage{xspace}

\graphicspath{{figs/}}

\newcommand{\sihi}{singlino/higgsino\xspace}

\newcommand{\lsim}
{\;\raisebox{-.3em}{$\stackrel{\displaystyle <}{\sim}$}\;}
\newcommand{\gsim}
{\;\raisebox{-.3em}{$\stackrel{\displaystyle >}{\sim}$}\;}

\newcommand\ReDiag{\mathop{%
  \raise .5pt\hbox{[}%
  \widetilde{\mathrm{Re}}%
  \raise .5pt\hbox{]}}}
\newcommand\ReOffDiag{\mathop{%
  \raise .5pt\hbox{$\llbracket$}%
  \widetilde{\mathrm{Re}}%
  \raise .5pt\hbox{$\rrbracket$}}}

\newcommand\Sl{\tilde l}

\newcommand\msl[1]{m_{\Sl_{#1}}}

\newcommand\gl{{\tilde g}}
\newcommand\mgl{m_\gl}

\newcommand\ino[1]{\tilde\chi_{#1}}

\newcommand\chapm[1]{\ino{#1}^\pm}

\newcommand\cha{\chapm}
\newcommand\mcha[1]{m_{\chapm{#1}}}

\newcommand\neu[1]{\ino{#1}^0}
\newcommand\mneu[1]{m_{\neu{#1}}}

\newcommand\refse[1]{Sect.~\ref{#1}}

\newcommand\citere[1]{Ref.~\cite{#1}}
\newcommand\citeres[1]{Refs.~\cite{#1}}

\newcommand{\CP}{{\cal CP}}
\newcommand{\cp}{{\CP}}

\newcommand{\tev}{\,\, \mathrm{TeV}}
\newcommand{\gev}{\,\, \mathrm{GeV}}

\newcommand{\Hz}{h_2}

\newcommand{\mHz}{m_{\Hz}}

\newcommand{\ETmiss}{\ensuremath{E_T\hspace{-4.0mm}\slash}\hspace{2.5mm}}

\newcommand\MO{\texttt{MicrOMEGAs}}

\newcommand\HB{\texttt{HiggsBounds}}
\newcommand\HS{\texttt{HiggsSignals}}
\newcommand\HT{\texttt{HiggsTools}}

\newcommand\pb{\ensuremath{\,\mbox{pb}}}

\newcommand{\sig}{\sigma}

\def\reffi#1{\mbox{Fig.~\ref{#1}}}

\def\De{\Delta}

\def\la{\lambda}
\def\ka{\kappa}

\def\gmin2{\ensuremath{(g-2)_\mu}}

\newcommand{\ssi}{\ensuremath{\sig_p^{\rm SI}}}

\newcommand{\Och}{\Omega_\chi h^2}

\definecolor{Orange}{named}{orange}
\definecolor{Purple}{named}{purple}
\definecolor{Lightblue}{cmyk}{0.9,0.1,0.1,0.3}
\definecolor{dgelborange}{cmyk}{0.,0.3,0.5, 0.}
\definecolor{Lila}{rgb}{0.5,0.,1}
\definecolor{Darkgreen}{rgb}{0.,.7,0.2}

\newcommand{\hto}[1]{{\color{Orange}  #1}}

\title{GUT-based NMSSM with Singlino Dark Matter:
Describing the ATLAS/CMS Excesses}

\author[a]{E.~Bagnaschi}
\author[b]{M.~Chakraborti}
\author*[c]{S.~Heinemeyer}
\author[d]{I.~Saha}

\affiliation[a]{INFN, Laboratori Nazionali di Frascati, Via E.\ Fermi 40,
00044 Frascati (RM), Italy}

\affiliation[b]{School of Physics and Astronomy, University of Southampton,
Southampton, SO17 1BJ,\\
United Kingdom (former address)}

\affiliation[c]{Instituto de F\'isica Te\'orica (UAM/CSIC), 
Universidad Aut\'onoma de Madrid,  Cantoblanco,\\
28049, Madrid, Spain}

\affiliation[d]{Department of Physics, Indian Institute of Technology Madras, Chennai 600036, India}

\emailAdd{emanuele.angelo.bagnaschi@lnf.infn.it}
\emailAdd{M.Chakraborti@soton.ac.uk}
\emailAdd{Sven.Heinemeyer@cern.ch}
\emailAdd{ipsita@iitm.ac.in}

\abstract{
One of the main goals of the ongoing LHC program is the search for
BSM physics, with EW SUSY partners still allowed with masses as low as
a few hundred GeV. 
Over the last years, searches for the ``golden channel'',
$pp \to \neu2 \cha1 \to \neu1 Z^{(*)} \, \neu1 W^{\pm (*)}$ show consistent
excesses between CMS and ATLAS in the 2~soft-lepton and 3~soft-lepton 
plus \ETmiss\ searches, favoring mass scales of
$\mneu2 \approx \mcha1 \gsim 200 \gev$ and
$\De m_{21} := \mneu2 - \mneu1 \approx 20 \gev$.
We analyze these excesses 
in the framework of the Next-to-Minimal Supersymmetric Standard Model
(NMSSM). We assume a singlino dominated lighest neutralino, $\neu1$,
as our Dark Matter (DM) candidate. The second and third lightest
neutralinos, $\neu{2,3}$ are higgsino like, with the higgsino mixing
parameter $\mu$ being smaller than the soft SUSY-breaking bino and
wino masses, $M_1$ and $M_2$. In particular, we assume the approximate
GUT relations $M_1 \sim M_2/2 \sim M_3/6$, yielding
a gluino mass $\mgl \sim M_3 \gsim 2 \tev$, in agreement with the LHC
search limits. The scalar masses are assumed to heavy and do not play a
role in our analysis. We demonstrate 
that this scenario is in agreement with all relevant
experimental constraints, comprising
the DM direct detection limits and the upper limit on the DM relic density, 
the LHC searches for SUSY particles and additional Higgs bosons,
as well as the LHC Higgs-boson rate measurements.
This constitutes the first explanation of the soft-lepton excesses in
a model with GUT relations among the soft SUSY-breaking 
parameters. 
}

\FullConference{19th International Conference on Topics in Astroparticle and Underground Physics (TAUP2025)\\
24–30 Aug 2025\\
Xichang, China \hfill {\tt IFT-UAM/CSIC-25-160}}

\begin{document}
\maketitle


\section{Introduction}
\label{sec:intro}

Within the (Next-to-)Minimal Supersymmetric Standard Model
((N)MSSM)~\cite{Ni1984,Ba1988,HaK85,GuH86}
(\cite{Maniatis:2009re,Ellwanger:2009dp})
the electroweak (EW) sector contains the 
SUSY partners of the SM leptons, the scalar leptons (sleptons), as
well as the charged (neutral) supersymmetric (SUSY) partners of the
charged (neutral) EW gauge bosons (gauginos) and Higgses (higgsinos).
The gauginos and higgsinos mix to form two charginos ($\cha{1,2}$)
and four (five) neutralinos ($\neu{1,2,3,4(,5)}$),
the so-called ``EWinos''. We assume the $\neu1$ to be a Dark Matter (DM)
candidate~\cite{Go1983,ElHaNaOlSr1984}. 
The gaugino/higgsino sector of the (N)MSSM contains
two soft SUSY-breaking parameters, $M_{1,2}$, and the the
SUSY-conserving parameter~$\mu$. In the NMSSM one has
additionally the paramters $\la$, $\ka$, $A_\ka$ and $A_\la$, see
\citere{BCHS} for our notation.

ATLAS and CMS are actively searching for the lighter EWinos,
$\neu{2,3}$ and $\cha1$, 
in final states containing two or three (possibly soft) leptons accompanied by
substantial missing transverse energy ($\ETmiss$)~\cite{ATLAS-SUSY,CMS-SUSY,Dicus:1983cb,Chamseddine:1983eg,Baer:1985at,Baer:1986vf,Baer:1986dv}. 
Interestingly, over the last years searches for the ``golden
channel'' (focusing for simplicity to the MSSM particle content), 
$pp \to \neu2 \cha1 \to \neu1 Z^{*} \, \neu1 W^{\pm *}$ show
consistent 
excesses between ATLAS and CMS in the 2~soft-lepton and 3~soft-lepton
plus missing energy~\cite{ATLAS:2019lng,CMS:2021edw}. This also holds for
the combined 2/3~soft-leptons plus
\ETmiss~\cite{ATLAS:2021moa,CMS:2024gyw,CMS:2025ttk}
searches assuming $\mneu2 \approx \mcha1$ for a mass difference of 
$\De m_{21} := \mneu2 - \mneu1 \approx 20 \gev$, referred
to as ``compressed spectra''. 

In \citere{CHS6}%
\footnote{
Other analyses of the soft-lepton excesses interpreted within the (N)MSSM framework can be found in \citeres{Ellwanger:2024vvs,Agin:2024yfs,Martin:2024pxx,Agin:2025vgn,Hammad:2025wst,Araz:2025bww}.
}%
~within the MSSM
it was found that the ``wino/bino DM'' scenario ($M_1 \lsim M_2 < \mu$)
yields a good description of the excesses.
On the other hand, it was concluded that
the ``higgsino DM'' scenario cannot ($|\mu| < |M_1|, M_2 $), 
cannot yield a sufficiently large $\De m_{21}$. This was
particularly due to the bounds from DD experiments, excluding the
range $\De m_{21} \gsim 10 \gev$.
The wino/bino DM scearnio requires $|M_1| \sim |M_2|$. On the other
hand, one of the most compelling arguments in favor of SUSY is the
unification of forces at the Grand Unified Theory (GUT)
scale. In simple GUT realizations, such as the
CMSSM, this yield the following mass pattern at the EW
scale: $M_1 \sim M_2/2 \sim M_3/6$, with the gluino mass $\mgl \sim M_3$.
LHC searches for colored particles have set a
limit of $\mgl \gsim 2 \tev$,
yielding $|M_1| \gsim 350 \gev$. Consequently, the
MSSM scenarios that can describe the soft-lepton excesses do not
follow the GUT-based mass pattern. 

In this paper, following \citere{BCHS}, we review a scenario that can
describe well the soft-lepton excesses, follows the GUT-based mass
patterns and is in 
agreement with the searches for gluinos: the NMSSM with a singlino
dominated LSP, with $200 \gev \approx |\mu| < M_1 \sim M_2/2 \sim M_3/6$, 
and $\mgl \gsim 2 \tev$. This parameter space corresponds to
higgsino-like $\neu{2,3}$ and $\cha1$, the ``\sihi'' scenario, 
which is furthermore in agreement with all limits
from DM measurements and searches.


\section {Set-up and constraints}
\label{sec:constraints}

For our notation, see \citere{BCHS}. 
We use the code {\tt NMSSMTools}~\cite{Ellwanger:2004xm,Ellwanger:2005dv,Domingo:2015qaa,NMSSMTOOLS-www}
as our spectrum generator. DM related quantities are calculated with
\MO{\tt-6.0}~\cite{Belanger:2001fz,Belanger:2006is,Belanger:2007zz,Belanger:2013oya,Belanger:2014vza,Barducci:2016pcb,Belanger:2018ccd,Belanger:2020gnr,Alguero:2023zol}  as included in {\tt NMSSMTools}.
Here we briefly list the experimental constraints that we apply
(if not mentioned otherwise, we use the constraints as
implemented into {\tt NMSSMTools}), for details see \citere{BCHS}:\\[-2.5em]

\begin{itemize}

\item Vacuum stability constraints\\[-2em]

\item Landau poles in the running couplings of the theory up to
  $M_{\mathrm{GUT}}$.\\[-1.5em]

\item
DM relic density constraints:
$\Omega_{\rm CDM} h^2 \; \le \; 0.122$~~\cite{Planck}.\\[-2em]

\item
DM direct detection (DD): spin-independent (SI) 
DM scattering cross-section ($\ssi$) limits from the PandaX-4T
experiment~\cite{PandaX:2024qfu}.\\[-2em] 

\item Constraints from LHC Higgs-boson rate measurements:
 tested with the code\\
\HS\,\texttt{v.3}~\cite{Bechtle:2013xfa,Bechtle:2014ewa,Bechtle:2020uwn,Bahl:2022igd}, 
which is included in the code \HT~\cite{Bahl:2022igd,HTnew}.\\[-2em]

\item Constraints from direct Higgs-boson searches at the LHC:
taken into account via\\ 
\HB\,\texttt{v.6}~\cite{Bechtle:2008jh,Bechtle:2011sb,Bechtle:2013wla,Bechtle:2015pma,Bechtle:2020pkv,Bahl:2022igd}, which is included in the public code \HT~\cite{Bahl:2022igd,HTnew}.\\[-2em]

\item Constraints from flavor physics.\\[-2em]

\item Searches for EWinos at LEP: $\mcha1 \gtrsim 100$ GeV, leading to
$|\mu| \gsim 100 \gev$, for our scenario.\\[-2em]

\item Searches for EWinos at the LHC:
In \citere{Agin:2024yfs} a reinterpretation of the original ATLAS 
data in the context of
a \sihi NMSSM scenario has been performed. We use their
exclusion lines as given in Fig.~8 of \citere{Agin:2024yfs} to exclude
low mass points from our data sample.\\[-2em]

\end{itemize}

We scan the following parameter space (again we refer to \citere{BCHS}
for our notation):\\[-2.0em]
\noindent
\begin{align}
  1000 \gev \leq 2 \times M_1 &= M_2 = M_3/3 \leq 2000 \gev \;, \quad
  150 \gev < | \mu | < 300 \gev \;, \notag\\
  \quad 1000 \gev &= \msl{L} = \msl{R} \;, 
  \quad A_e = A_{\mu} = A_{\tau} = 0\;, \notag\\
  5000 \gev &= M_{\tilde{Q}_{1,2}} = M_{\tilde{u}_R, \tilde{d}_R, \tilde{c}_R, \tilde{s}_R} \;, \quad A_{u,d,c,s} = 0\;, \notag \\
  3000 \gev = M_{\tilde{Q}_3} &= M_{\tilde{t}_R, \tilde{b}_R} \;, \quad A_{b} = 0 \;, \quad 5000 \gev < A_t < 8000 \gev\;, \notag \\
  -0.7 &\le \kappa < 0.7 \;, \quad 0 < \lambda < 5 \;, \quad 0.5 \le \tan\beta \le 50\;, \notag \\
  5 \gev &\le \left| A_{\lambda} \right| \le 5000 \gev \;, \quad -1000 \gev \le A_{\kappa} \le 1000 \gev\,.
  \label{sihip}
\end{align}


\section{Results}
\label{sec:results}

Our result plots (below) show different colors, depening on which
experimental constraint, see \refse{sec:constraints}, yields the
exclusion.
The two most important colors to understand the plots are:\\[-2em]
\begin{itemize}
\item non-red: all scan points that were saved at the sampling level.\\[-2em]

\item red: points that pass all constraints, particularly
searches for $pp \to \neu{2,3} \cha1$ with small $\De m_{21}$.\\[-2em]
  
\end{itemize}

\begin{figure}[htb!]
\centering
\centering\includegraphics[width=0.6\textwidth]{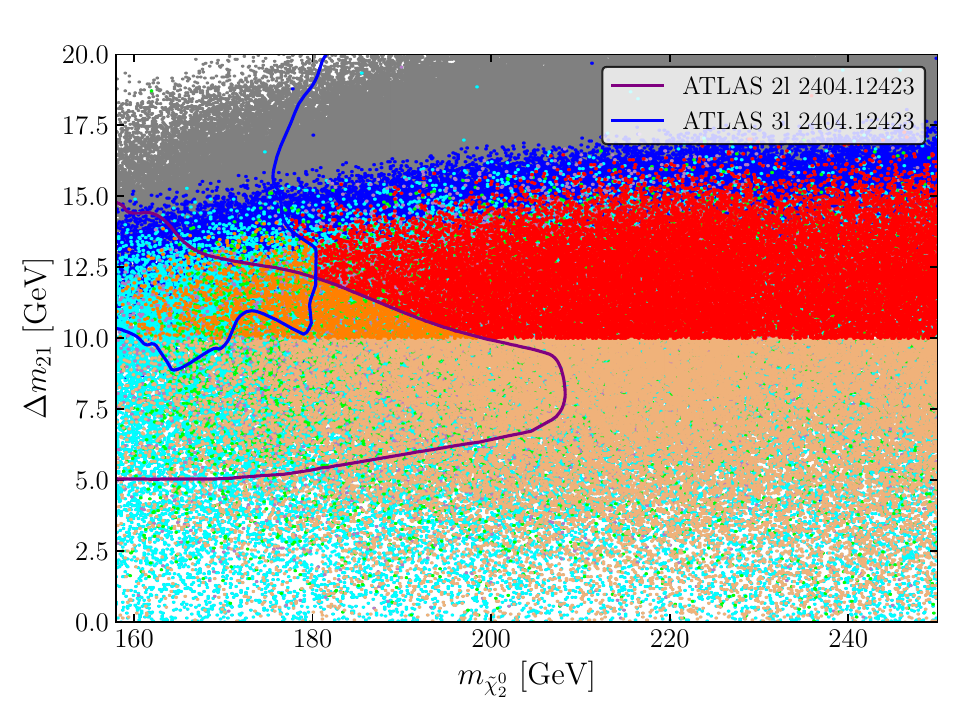}\\
\includegraphics[width=0.48\textwidth]{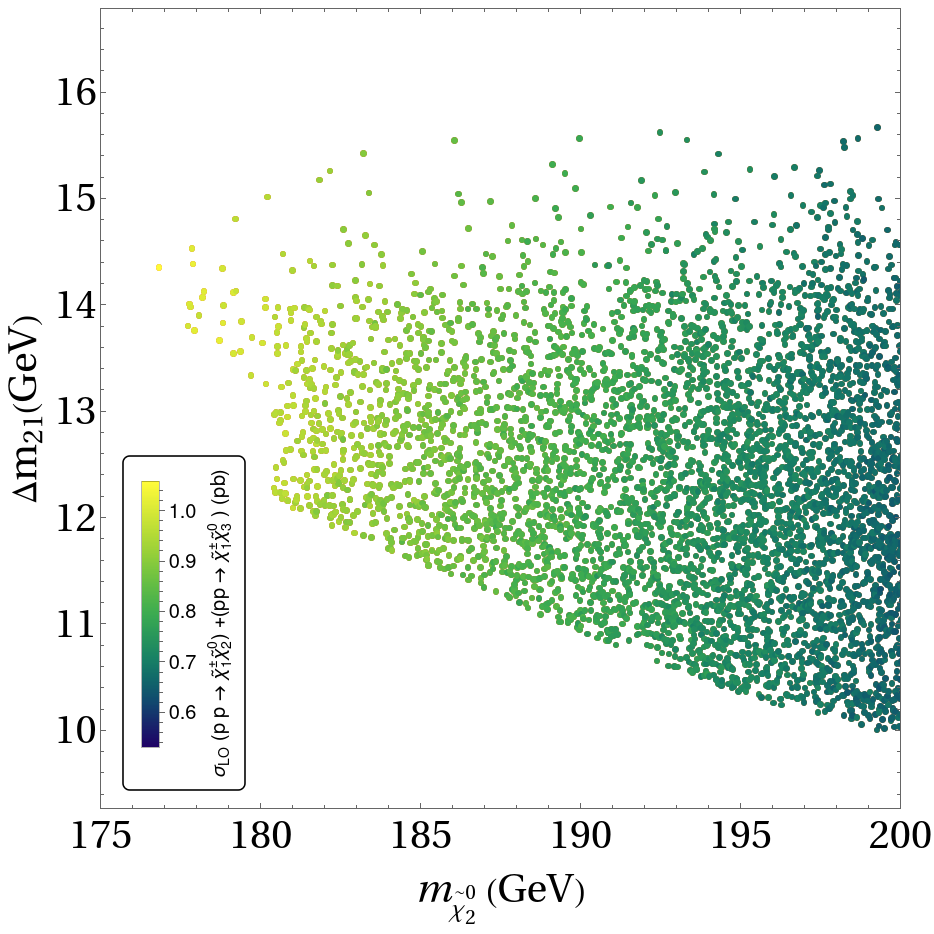}
\includegraphics[width=0.48\textwidth]{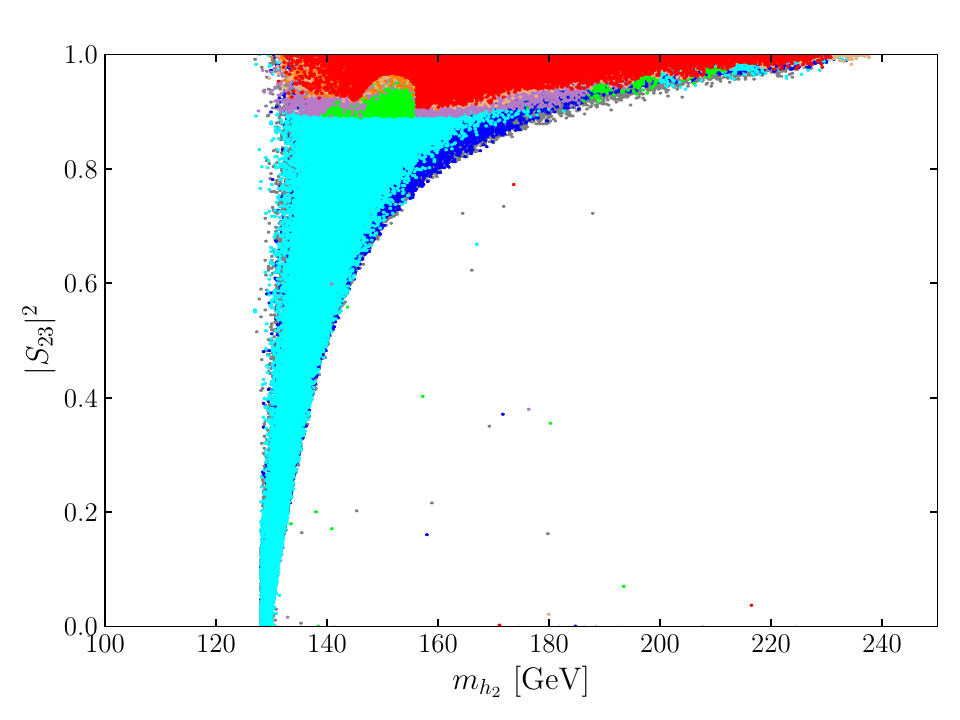}\\
\includegraphics[width=0.48\textwidth]{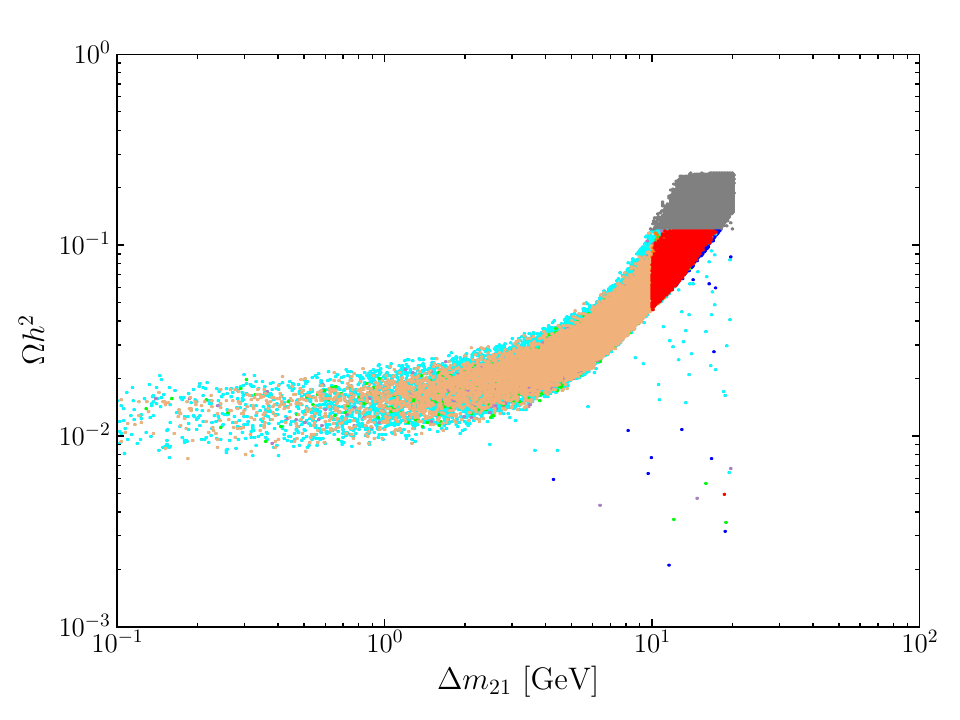}
\includegraphics[width=0.48\textwidth]{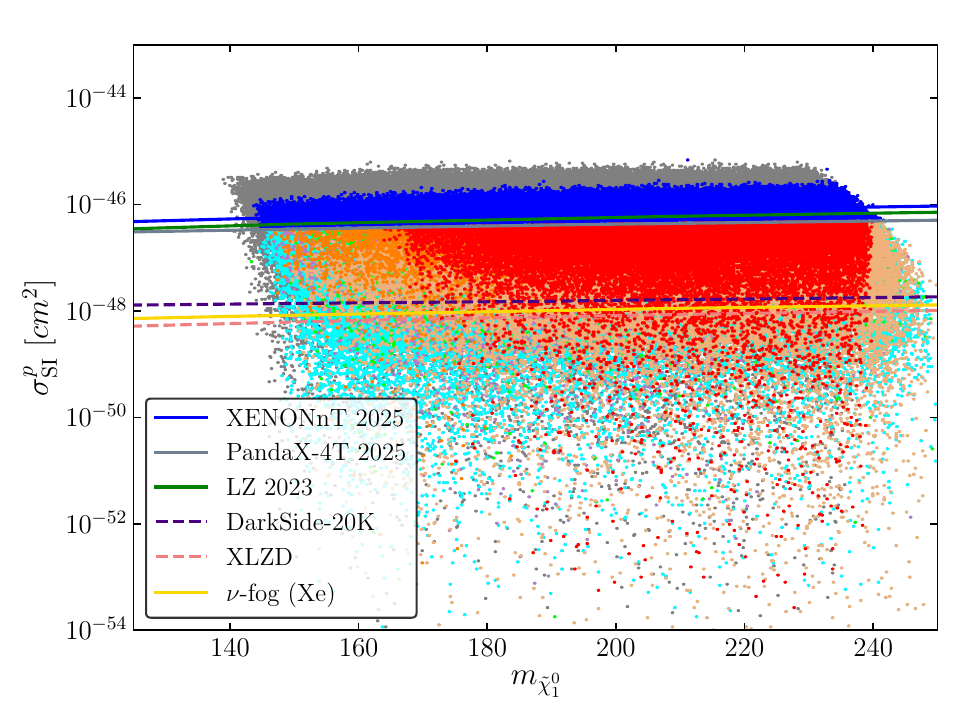}
\caption{The results of our parameter scan in the \sihi scenario.
Upper plot: $\mneu2$--$\De m_{21}$ plane;
middle left: predictions of $\sig(pp \to \neu2\cha1) + \sig(pp \to \neu3\cha1)$
in the $\mneu2$--$\De m_{21}$ plane;
middle right: $|S_{23}|^2$--$m_{h_2}$ plane; 
lower left: $\Och$--$\De m_{21}$ plane;
lower right: $\mneu1$--$\ssi$ plane (see text).}
\label{fig:res}
\end{figure}

In \reffi{fig:res} we review the most important findings of our
results~\cite{BCHS}. 
We start in the upper row of \reffi{fig:res}, where we show
the $\mneu2$--$\De m_{21}$ plane, overlaid with the experimental 95\%
CL exclusion 
bounds as recasted in \citere{Agin:2024yfs}. The blue line indicates
the ATLAS 3~soft-lepton limits~\cite{ATLAS:2021moa}, whereas the purple
line indicates the ATLAS 2~soft-lepton limits~\hto{\cite{ATLAS:2019lng}}.%
\footnote{The recasting performed in \citere{Agin:2024yfs} was performed only for the ATLAS limtis.
However, it has been discussed in the literature, see e.g.\ \citere{CHS6}, that the ATLAS and CMS limits
effectively show the same excesses in the 3~and 2~soft-lepton channels.}
~One can observe that
our red points, i.e.\ the ones passing all constraints stretch out
from the 95\% CL exlusion lines up to $\mneu2 = 250 \gev$, where our
scan stopped. $\De m_{21} := \mneu2 - \mneu1$ is found between $10 \gev$
(where we cut the ``preferred parameter space'') 
and $\sim 16 \gev$, with the values closest to the exclusion bounds of
$12 \gev \lsim \De m_{21} \lsim 14 \gev$. This plot clearly demonstrates
that the \sihi scenario in the NMSSM yields a very good description of
the soft-lepton excesses observed at ATLAS and CMS, while being in
agreement with all other theoretical and experimental constraints.
The fact that $\De m_{21}$ comes out somewhat smaller in the NMSSM as
compared to the preferred $\sim 20 \gev$ in the MSSM (see,
e.g., \citere{CHS6}) can be attributed to the fact that two production
channels are open in the NMSSM, $pp \to \neu{2,3}\cha1$, as compared
to only one in the MSSM, where by definition we have $\mneu3 > \mneu2$.

In the middle left plot of \reffi{fig:res} we show our predictions for
$\sig(pp \to \neu2\cha1) + \sig(pp \to \neu3\cha1)$ at the LHC at
$\sqrt{s} = 13 \tev$ in the $\mneu2$--$\De m_{21}$ plane.
We computed the leading-order (LO) production cross-section using
{\tt MadGraph (MG5aMC)-v3.6.3}~\cite{Alwall:2014hca} for which we  
incorporated the NMSSM model file from the
{\tt FeynRules}~\cite{Alloul:2013bka,Conte:2016zjp} database. 
To obtain the NLO cross-section, we applied the MSSM k-factor given by 
{\tt Resummino}~\cite{resummino,Bozzi:2006fw,Bozzi:2007qr,Debove:2009ia,Debove:2010kf} to the LO production cross-section. 
The size of the cross section for each point is indicated by the color
scale. The smallest EWino masses yield production cross sections of
approximately $\sim 1.3 \pb$, going down to $\sim 0.7 \pb$ 
for $\mneu2 \sim 200 \gev$ (where our plot ends).
These values are roughly consistent with the cross sections required to
reproduce the observed excess in events, within the associated  
uncertainties, see e.g.\ the discussion in \citere{CHS6}.
This defines a clear target for Run~3 of the LHC and future collider stages. 

In the middle right plot of \reffi{fig:res} we review the results for
the second lightest $\cp$-even Higgs, $h_2$ in the $\mHz$--$|S_{23}|^2$ plane,
i.e.\ its mass vs.\ its singlet component.
We find that the mass is restricted to be
$125 \gev \lsim \mHz \lsim 225 \gev$, with a singlet component of 90\%
or larger. This indicates that the soft-lepton excesses at ATLAS and
CMS, when interpreted in the NMSSM, predict a second light
Higgs-boson, which, however, largely decouples from the SM
particles. Nevertheless, this light Higgs-boson might be an interesting
target for future HL-LHC direct searches. We leave a detailed
phenomenologcial analysis for future work.

Now we turn to the predictions for DM in our \sihi
scenario. In the lower left plot of \reffi{fig:res} we review our scan
results in the $\De m_{21}$--$\Och$ plane. A clear correlation over the whole
scanned parameter space is visible, with lower DM relic densities
reached for smaller $\De m_{21}$. Turning this around, larger $\De m_{21}$,
as favored by the soft-lepton excesses leads naturally to a (possible)
saturation of the DM relic density. This is in stark contrast to the higgsino
scenarios in the MSSM, see \citere{CHS6}, where always a strongly
underabundant DM relic density was found.
The prospects for spin-independent ($\ssi$) direct DM
searches in our \sihi scenario are presented 
in the lower right plot of \reffi{fig:res}, where we show the scan
results in the $\mneu1$--$\ssi$ plane.and indicate with solid blue,
black and gray lines the current bounds from
XENON-nT~\cite{XENON:2025vwd}, PandaX-4T~\cite{PandaX:2024qfu} and
LZ~\cite{LZ-new}, respectively. As discussed in \refse{sec:constraints}
and visible in the plot, we applied the limits from
PandaX-4T~\cite{PandaX:2024qfu}. Having applied the more recent
published limits from LZ~\cite{LZ:2024zvo} would cut away the points
with the highest DD cross sections. However, we do not expect a major
relevant impact on our general findings. As dashed purple and red lines
we indicate the anticipated limits from DarkSide-20K
XLZD, repsectively. The $\nu$-fog (for Xenon) is shown as yellow solid line.
While the future experiments can cover a large part of the preferred
parameter space, the NMSSM interpretation of the 2- and 3-soft lepton
excesses are also compatible with points prediciting a DD cross section
below the $\nu$-fog limit, i.e.\ not observable with currently used
experimental set-ups. We have tested that the points below the $\nu$-fog
limit correspond to somewhat smaller $\De m_{21}$, not 
exceeding $13 \gev$ for the smallest allowed $\mneu2$ values.

\smallskip
Overall, the \sihi\ DM scenario offers various prospects for experimental
tests {\it complementary} to the direct production of light higgsinos.

\subsection*{Acknowledgments}
I.S.~acknowledges support from SERB-MATRICS, ANRF, India, under grant
no. MTR/2023/ 000715. 
The work of S.H.\ has received financial support from the
grant PID2019-110058GB-C21 funded by
MCIN/AEI/10.13039/501100011033 and by ``ERDF A way of making Europe", 
and in part by the grant IFT Centro de Excelencia Severo Ochoa
CEX2020-001007-S funded by MCIN/AEI/10.13039/501100011033. 
S.H.\ also acknowledges support from Grant PID2022-142545NB-C21 funded by
MCIN/AEI/10.13039/501100011033/ FEDER, UE. EB would like to thank the
LNF INFN Data Cloud Team 
and the CERN-IT department for the use of their computational resources.


\newcommand\jnl[1]{\textit{\frenchspacing #1}}
\newcommand\vol[1]{\textbf{#1}}



\begin{thebibliography}{999}

%
%
%
%
%

\bibitem{Ni1984}
H.~Nilles, 
\jnl{Phys. Rept.} \vol{110} (1984) 1.

\bibitem{Ba1988}
R.~Barbieri, 
\jnl{Riv. Nuovo Cim.} \vol{11} (1988) 1. 

\bibitem{HaK85}
H.~Haber, G.~Kane,
\jnl{Phys. Rept.} \vol{117} (1985) 75.

\bibitem{GuH86}
J.~Gunion, H.~Haber,
\jnl{Nucl. Phys.} \vol{B 272} (1986) 1.

\bibitem{Maniatis:2009re}
M.~Maniatis,
\jnl{Int. J. Mod. Phys.} \vol{A 25} (2010) 3505
[arXiv:0906.0777 [hep-ph]].

\bibitem{Ellwanger:2009dp} 
Ulrich Ellwanger, Cyril Hugonie, and A.M.Teixeira,
\jnl{Phys.Rept.} \vol{496} (2010) 1
[arXiv:0910.1785 [hep-ph]].

\bibitem{Go1983} 
H.~Goldberg,
\jnl{Phys. Rev. Lett.} \vol{50} (1983) 1419.

\bibitem{ElHaNaOlSr1984}
J.~Ellis, J.~Hagelin, D.~Nanopoulos, K.~Olive, M.~Srednicki,
\jnl{Nucl. Phys.} \vol{B 238} (1984) 453.

\bibitem{BCHS}
E.~Bagnaschi, M.~Chakraborti, S.~Heinemeyer and I.~Saha, 
arXiv:2512.16783 [hep-ph].

\bibitem{ATLAS-SUSY}
See: {\tt https://twiki.cern.ch/twiki/bin/view/AtlasPublic/\\SupersymmetryPublicResults}~.

\bibitem{CMS-SUSY}
See: {\tt https://twiki.cern.ch/twiki/bin/view/CMSPublic/PhysicsResultsSUS}~.

\bibitem{Dicus:1983cb}
D.~A.~Dicus, S.~Nandi and X.~Tata,
\jnl{Phys. Lett.} \vol{B 129} (1983), 451
[erratum: \jnl{Phys. Lett.} \vol{B 145} (1984), 448].

\bibitem{Chamseddine:1983eg}
A.~H.~Chamseddine, P.~Nath and R.~L.~Arnowitt,
\jnl{Phys. Lett.} \vol{B 129} (1983), 445
[erratum: \jnl{Phys. Lett.} \vol{B 132} (1983), 467].

\bibitem{Baer:1985at}
H.~Baer and X.~Tata,
\jnl{Phys. Lett.} \vol{B 155} (1985), 278-283.

\bibitem{Baer:1986vf}
H.~Baer, K.~Hagiwara and X.~Tata,
\jnl{Phys. Rev.} \vol{D 35} (1987), 1598.

\bibitem{Baer:1986dv}
H.~Baer, K.~Hagiwara and X.~Tata,
\jnl{Phys. Rev. Lett.} \vol{57} (1986), 294.

\bibitem{ATLAS:2019lng}
G.~Aad \textit{et al.} [ATLAS],
\jnl{Phys. Rev.} \vol{D 101} (2020) no.5, 052005
[arXiv:1911.12606 [hep-ex]].

\bibitem{CMS:2021edw}
A.~Tumasyan \textit{et al.} [CMS],
\jnl{JHEP} \vol{04} (2022), 091
[arXiv:2111.06296 [hep-ex]].

\bibitem{ATLAS:2021moa}
G.~Aad \textit{et al.} [ATLAS],
\jnl{Eur. Phys. J.} \vol{C 81} (2021) no.12, 1118
[arXiv:2106.01676 [hep-ex]].

\bibitem{CMS:2024gyw}
A.~Hayrapetyan \textit{et al.} [CMS],
arXiv:2402.01888 [hep-ex];

\bibitem{CMS:2025ttk}
V.~Chekhovsky \textit{et al.} [CMS],
[arXiv:2508.13900 [hep-ex]].

\bibitem{CHS6}
M.~Chakraborti, S.~Heinemeyer and I.~Saha,
\jnl{Eur. Phys. J.} \vol{C 84} (2024) no.8, 812
[arXiv:2403.14759 [hep-ph]].

\bibitem{Ellwanger:2024vvs}
U.~Ellwanger, C.~Hugonie, S.~F.~King and S.~Moretti,
\jnl{Eur. Phys. J.} \vol{C 84}, no.8, 788 (2024)
[arXiv:2404.19338 [hep-ph]].

\bibitem{Agin:2024yfs}
D.~Agin, B.~Fuks, M.~D.~Goodsell and T.~Murphy,
\jnl{Eur. Phys. J.} \vol{C 84}, no.11, 1218 (2024)
[arXiv:2404.12423 [hep-ph]].

\bibitem{Martin:2024pxx}
S.~P.~Martin,
\jnl{Phys. Rev.} \vol{D 109}, no.9, 095045 (2024)
[arXiv:2403.19598 [hep-ph]].

\bibitem{Agin:2025vgn}
D.~Agin, B.~Fuks, M.~D.~Goodsell and T.~Murphy,
\jnl{Eur. Phys. J.} \vol{C 85}, no.10, 1145 (2025)
[arXiv:2506.21676 [hep-ph]].

\bibitem{Hammad:2025wst}
A.~Hammad, R.~Ramos, A.~Chakraborty, P.~Ko and S.~Moretti,
[arXiv:2508.13912 [hep-ph]].

\bibitem{Araz:2025bww}
J.~Y.~Araz, B.~Fuks, M.~D.~Goodsell and T.~Murphy,
[arXiv:2507.08927 [hep-ph]].

\bibitem{Ellwanger:2004xm}
U.~Ellwanger, J.~Gunion, C.~Hugonie,
\jnl{JHEP} \vol{02} (2005) 066
[arXiv:hep-ph/0406215].

\bibitem{Ellwanger:2005dv}
U.~Ellwanger, C.~Hugonie,
\jnl{Comput.\,Phys.\,Commun.} \vol{175} (2006) 290,
[arXiv:hep-ph/0508022].

\bibitem{Domingo:2015qaa}
F.~Domingo,
\jnl{JHEP} \vol{06} (2015) 052,
[arXiv:1503.07087 [hep-ph]].

\bibitem{NMSSMTOOLS-www}
See: \texttt{https://www.lupm.univ-montp2.fr/users/nmssm/}.

\bibitem{Belanger:2001fz}
G.~Belanger, F.~Boudjema, A.~Pukhov and A.~Semenov,
\jnl{Comput. Phys. Commun.} \vol{149} (2002), 103-120
[arXiv:hep-ph/0112278 [hep-ph]].

\bibitem{Belanger:2006is}
G.~Belanger, F.~Boudjema, A.~Pukhov and A.~Semenov,
\jnl{Comput. Phys. Commun.} \vol{176} (2007), 367-382
[arXiv:hep-ph/0607059 [hep-ph]].

\bibitem{Belanger:2007zz}
G.~Belanger, F.~Boudjema, A.~Pukhov and A.~Semenov,
\jnl{Comput. Phys. Commun.} \vol{177} (2007), 894-895.

\bibitem{Belanger:2013oya}
  G.~Belanger, F.~Boudjema, A.~Pukhov and A.~Semenov,
  arXiv:1305.0237 [hep-ph].

\bibitem{Belanger:2014vza}
G.~B{\'e}langer, F.~Boudjema, A.~Pukhov and A.~Semenov,
\jnl{Comput. Phys. Commun.} \vol{192} (2015), 322-329
[arXiv:1407.6129 [hep-ph]].

\bibitem{Barducci:2016pcb}
D.~Barducci, G.~Belanger, J.~Bernon, F.~Boudjema, J.~Da Silva, S.~Kraml, U.~Laa and A.~Pukhov,
\jnl{Comput. Phys. Commun.} \vol{222} (2018), 327-338
[arXiv:1606.03834 [hep-ph]].

\bibitem{Belanger:2018ccd}
G.~B{\'e}langer, F.~Boudjema, A.~Goudelis, A.~Pukhov and B.~Zaldivar,
\jnl{Comput. Phys. Commun.} \vol{231} (2018), 173-186
[arXiv:1801.03509 [hep-ph]].

\bibitem{Belanger:2020gnr}
G.~Belanger, A.~Mjallal and A.~Pukhov,
\jnl{Eur. Phys. J.} \vol{C 81} (2021) no.3, 239
[arXiv:2003.08621 [hep-ph]].

\bibitem{Alguero:2023zol}
G.~Alguero, G.~Belanger, F.~Boudjema, S.~Chakraborti, A.~Goudelis, S.~Kraml, A.~Mjallal and A.~Pukhov,
\jnl{Comput. Phys. Commun.} \vol{299} (2024), 109133
[arXiv:2312.14894 [hep-ph]].

\bibitem{Planck}
 N.~Aghanim {\it et al.} [Planck Collaboration],
\jnl{Astron. Astrophys.} \vol{641} (2020), A6
[erratum: \jnl{Astron. Astrophys.} \vol{652} (2021), C4]
[arXiv:1807.06209 [astro-ph.CO]].

\bibitem{LZ-new}
J.~Aalbers \textit{et al.} [LZ],
\jnl{Phys. Rev. Lett.} \vol{131} (2023) no.4, 041002
[arXiv:2207.03764 [hep-ex]].

\bibitem{Bechtle:2013xfa}
  P.~Bechtle, S.~Heinemeyer, O.~St{\aa}l, T.~Stefaniak and G.~Weiglein,
  \jnl{Eur.\ Phys.\ J.} \vol{C 74} (2014) no.2,  2711
  [arXiv:1305.1933 [hep-ph]].

\bibitem{Bechtle:2014ewa}
  P.~Bechtle, S.~Heinemeyer, O.~St{\aa}l, T.~Stefaniak and G.~Weiglein,
  \jnl{JHEP} \vol{1411} (2014) 039
  [arXiv:1403.1582 [hep-ph]].

\bibitem{Bechtle:2020uwn}
P.~Bechtle, S.~Heinemeyer, T.~Klingl, T.~Stefaniak, G.~Weiglein and
J.~Wittbrodt, 
\jnl{Eur. Phys. J.} \vol{C 81} (2021) no.2, 145
[arXiv:2012.09197 [hep-ph]].

\bibitem{Bahl:2022igd}
H.~Bahl, T.~Biek\"otter, S.~Heinemeyer, C.~Li, S.~Paasch, G.~Weiglein
and J.~Wittbrodt, 
\jnl{Comput. Phys. Commun.} \vol{291} (2023), 108803
[arXiv:2210.09332 [hep-ph]]. 

\bibitem{HTnew}
H.~Bahl, T.~Biek\"otter, S.~Heinemeyer, K.~Radchenko and G.~Weiglein, 
{\tt IFT–UAM/CSIC-25-072} (to appear).

\bibitem{Bechtle:2008jh}
  P.~Bechtle, O.~Brein, S.~Heinemeyer, G.~Weiglein and K.~E.~Williams,
  \jnl{Comput.\ Phys.\ Commun.}  \vol{181} (2010) 138
  [arXiv:0811.4169 [hep-ph]].

\bibitem{Bechtle:2011sb}
  P.~Bechtle, O.~Brein, S.~Heinemeyer, G.~Weiglein and K.~E.~Williams,
  \jnl{Comput.\ Phys.\ Commun.}  \vol{182} (2011) 2605
  [arXiv:1102.1898 [hep-ph]].

\bibitem{Bechtle:2013wla}
  P.~Bechtle, O.~Brein, S.~Heinemeyer, O.~St{\aa}l, T.~Stefaniak,
  G.~Weiglein and K.~E.~Williams, 
  \jnl{Eur.\ Phys.\ J.} \vol{C 74} (2014) no.3,  2693
  [arXiv:1311.0055 [hep-ph]].

\bibitem{Bechtle:2015pma}
  P.~Bechtle, S.~Heinemeyer, O.~St{\aa}l, T.~Stefaniak and G.~Weiglein,
  \jnl{Eur.\ Phys.\ J.} \vol{C 75} (2015) no.9,  421
  [arXiv:1507.06706 [hep-ph]].

\bibitem{Bechtle:2020pkv}
P.~Bechtle, D.~Dercks, S.~Heinemeyer, T.~Klingl, T.~Stefaniak,
G.~Weiglein and J.~Wittbrodt,
\jnl{Eur. Phys. J.} \vol{C 80} (2020) no.12, 1211
[arXiv:2006.06007 [hep-ph]].

\bibitem{Alwall:2014hca}
J.~Alwall, R.~Frederix, S.~Frixione, V.~Hirschi, F.~Maltoni, O.~Mattelaer, H.~S.~Shao, T.~Stelzer, P.~Torrielli and M.~Zaro,
\jnl{JHEP} \vol{07}, (2014) 079
[arXiv:1405.0301 [hep-ph]].

\bibitem{Alloul:2013bka}
A.~Alloul, N.~D.~Christensen, C.~Degrande, C.~Duhr and B.~Fuks,
\jnl{Comput. Phys. Commun.} \vol{185}, (2014) 2250-2300 
[arXiv:1310.1921 [hep-ph]].

\bibitem{Conte:2016zjp}
E.~Conte, B.~Fuks, J.~Guo, J.~Li and A.~G.~Williams,
\jnl{JHEP} \vol{05}, (2016) 100
[arXiv:1604.05394 [hep-ph]].

\bibitem{resummino}
B.~Fuks, M.~Klasen, D.~R.~Lamprea and M.~Rothering,
\jnl{Eur. Phys. J.} \vol{C 73} (2013), 2480
[arXiv:1304.0790 [hep-ph]].

\bibitem{Bozzi:2006fw}
G.~Bozzi, B.~Fuks and M.~Klasen,
\jnl{Phys. Rev.} \vol{D 74} (2006), 015001
[arXiv:hep-ph/0603074 [hep-ph]].

\bibitem{Bozzi:2007qr}
G.~Bozzi, B.~Fuks and M.~Klasen,
\jnl{Nucl. Phys.} \vol{B 777} (2007), 157-181
[arXiv:hep-ph/0701202 [hep-ph]].

\bibitem{Debove:2009ia}
J.~Debove, B.~Fuks and M.~Klasen,
\jnl{Phys. Lett.} \vol{B 688} (2010), 208-211
[arXiv:0907.1105 [hep-ph]].

\bibitem{Debove:2010kf}
J.~Debove, B.~Fuks and M.~Klasen,
\jnl{Nucl. Phys.} \vol{B 842} (2011), 51-85
[arXiv:1005.2909 [hep-ph]].

\bibitem{XENON:2025vwd}
E.~Aprile \textit{et al.} [XENON],
\jnl{Phys. Rev. Lett.} \vol{135} (2025) no.22, 221003

\bibitem{PandaX:2024qfu}
Z.~Bo \textit{et al.} [PandaX],
\jnl{Phys. Rev. Lett.} \vol{134} (2025) no.1, 011805
[arXiv:2408.00664 [hep-ex]].

\bibitem{LZ:2024zvo}
J.~Aalbers \textit{et al.} [LZ],
\jnl{Phys. Rev. Lett.} \vol{135} (2025) no.1, 011802
[arXiv:2410.17036 [hep-ex]].


  
\end{thebibliography}
\end{document}